\documentclass[fleqn,twoside]{article}
\usepackage{espcrc2}

\usepackage{graphicx}

\newcommand{\psib}{\ensuremath{\overline{\psi}}}

\newcommand{\AmS}{{\protect\the\textfont2
  A\kern-.1667em\lower.5ex\hbox{M}\kern-.125emS}}

\title{A Two-Dimensional Lattice Model with Exact Supersymmetry}

\author{S. Catterall\address{Department of Physics, Syracuse University,
Syracuse, NY 13244} 
\thanks{Corresponding author: smc@physics.syr.edu},S. Karamov\addressmark
}
       
\begin{document}

\begin{abstract}
Starting from a simple discrete model which exhibits a
supersymmetric invariance we construct a local, interacting, two-dimensional
Euclidean lattice theory which also admits
an exact supersymmetry. This model is shown to correspond to the Wess-Zumino model
with extended $N=2$ supersymmetry in the continuum. 
We have performed dynamical fermion simulations to check the
spectrum and supersymmetric Ward identities and find good agreement
with theory.
\vspace{1pc}
\end{abstract}

\maketitle

\section{Introduction}
Supersymmetry is thought to be a crucial ingredient in any theory which  
attempts to unify the separate interactions contained in the standard model  
of particle physics. On rather general grounds we
expect that supersymmetry must be broken non-perturbatively
in any such theory. This has led to attempts to formulate
such theories on lattices \cite{general}. 

However, most lattice models break supersymmetry
explicitly and lead generically to the
appearance of relevant, SUSY
violating interactions in the lattice effective action. The
continuum limit will not then correspond to a supersymmetric theory
without fine tuning.

In this talk we show that models exist which may be discretized in
such a way as to preserve a subset of the continuum SUSY transformations
exactly. Such models may evade this fine tuning problem.

\section{Simple Model}
Consider a set of discrete, real, commuting fields $x_i$ and 
real, anticommuting  fields $\psi_i,\psib_i$ where $i=1\ldots K$
with action
\begin{displaymath}
S=\frac{1}{2}N_i^2(x)+\psib_i\frac{\partial N_i}{\partial x_j}\psi_j
\end{displaymath}
which admits a `supersymmetry'
\begin{eqnarray*}
\delta x_i &=& \psi_i\xi \nonumber\\
\delta \psib_i &=& N_i\xi \nonumber \\
\delta \psi_i &=& 0 \nonumber
\label{invariance}
\end{eqnarray*}
Corresponding to this symmetry the quantum
theory exhibits Ward identities
eg.
\begin{displaymath}
\left< \psi_i\psib_j\right>+\left< \overline{N}_i x_j\right>=0
\end{displaymath}
If we choose the fields $x,\psi,\psib$ to lie on a spatial
lattice equipped with periodic boundary conditions and take
the fermion matrix $M_{ij}=\frac{\partial N_i}{\partial x_j}$ to be of
the form
\[M_{ij}=D^S_{ij}+P^{\prime\prime}_{ij}(x)\]
we recover a Euclidean lattice version of SUSY QM
\begin{displaymath}
S=\frac{1}{2}(D^S_{ij}
x_j+P^\prime_i)^2+\psib_i(D^S_{ij}+P^{\prime\prime}_{ij})\psi_j
\end{displaymath}
Notice that 
if we include a Wilson term in
$P^{\prime\prime}$ we can eliminate
double modes in both fermionic and bosonic sectors. Notice also that
the lattice action contains a term which
behaves as a total derivative in the naive continuum limit. However,
on the lattice it is non-zero for an interacting theory and
its inclusion is necessary if the theory is to possess an exact
SUSY invariance.

\section{Mean Action}

\begin{figure}[htb]
\caption{Massgaps vs $a$ at $m=10$, $g=100$}
\label{fig1}
\includegraphics[width=6.0cm]{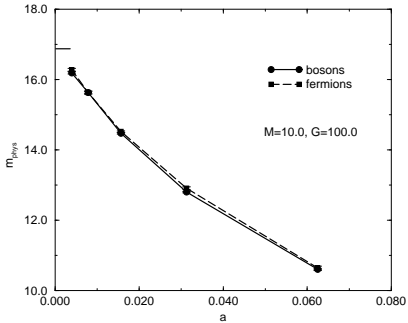}
\end{figure}

For $P^\prime=mx+gx^Q$  
we can show using a simple scaling argument
and a supersymmetric Ward identity
that the mean action $<S>=K$, the number of
degrees of freedom,
which we recognize as a
Euclidean analog of the vanishing of the
vacuum energy $E_{\rm vac}=0$ in a supersymmetric theory. 
Notice also that
S is approximately invariant under 2nd SUSY
\begin{eqnarray*}
\delta^\prime x_i &=& \psib_i\xi \nonumber \\
\delta^\prime \psi_i &=& (D_{ij}x_j-P^\prime_i)\xi \nonumber \\
\delta^\prime \psib_i &=& 0 \nonumber
\end{eqnarray*}
where
\begin{displaymath}
\delta^\prime S=\delta^\prime\left(2D_{ij}^Sx_jP^\prime_i\right)
\end{displaymath}
Since QM is a finite theory we hence
expect the continuum model to have the full $N=2$ SUSY. Fig~\ref{fig1}
shows a plot of the massgap in the model where
$P^\prime=mx+gx^3$ for $m=10$ and $g=100$, as a function of the
lattice spacing which clearly exhibits the degeneracy between bosonic
and fermionic degrees of freedom at finite lattice spacing. 
\section{Wess Zumino Model}
Imagine promoting the indices $i\to \left(i,\alpha\right)$ where
$i$ labels the spacetime point and
$\alpha=1,2$ a spinor component for a theory in two dimensions.
Notice immediately that the target theory will contain two
real scalars. 
We take a fermion matrix of the form
\begin{displaymath}
M_{ij}^{\alpha\beta}=\gamma^\mu_{\alpha\beta}D^\mu_{ij}+
A_{ij}\delta_{\alpha\beta}+B_{ij}i\gamma^5_{\alpha\beta}
\end{displaymath}
We find that
this matrix can only be obtained by differentiating some
vector field if the bosons possess a complex structure
$\phi_i=x^1_i+ix^2_i$ and
\begin{displaymath}
S=\frac{1}{2}\overline{\eta}^{(1)}\eta^{(1)}+\psib M\psi
\end{displaymath}
where
\begin{displaymath}
\eta_i^{(1)}=D_z\phi_i+W^\prime_i(\phi)
\end{displaymath}
and
$W(\phi)$ is an arbitrary analytic
function.

We can also write down three other complex fields which yield
the same continuum bosonic action
\begin{eqnarray*}
\eta^{(2)}&=&D_z\overline{\phi}-W^\prime(\phi)\nonumber\\
\eta^{(3)}&=&D_z\phi-W^\prime(\overline{\phi})\nonumber\\
\eta^{(4)}&=&D_z\phi+W^\prime(\overline{\phi})
\end{eqnarray*}
All these lead to same value for ${\rm det}M$ and generate
approximate symmetries on the lattice
$\delta S\sim ga^2$ eg.
\begin{eqnarray}
\delta_2 x_i &=& i\gamma_5^{\alpha\beta}\psi_i^{\beta}\xi \nonumber \\
\delta_2 \psib_i^\alpha &=& i\gamma_5^{\alpha\beta}\overline{N}_i^\beta\xi \nonumber \\
\delta_2 \psi_i^\alpha &=& 0 \nonumber
\end{eqnarray}
with
\begin{displaymath}
\eta^{(2)}_i=(\overline{N}^1_i+i\overline{N}^2_i)
\end{displaymath}
We can also derive Ward identities corresponding to all these
exact and approximate supersymmetries.

\section{Simulations}
\begin{figure}[htb]
\caption{1st Ward identity}
\label{fig3}
\includegraphics[width=6.0cm]{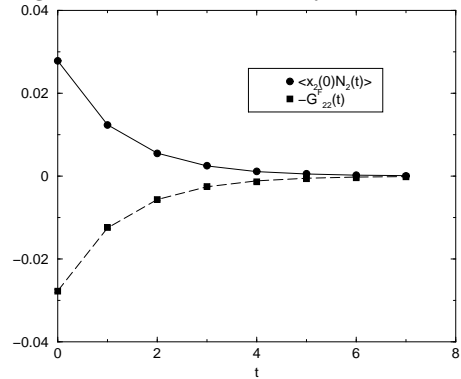}
\end{figure}

We have checked these conclusions by explicit dynamical
fermion simulations for the case $W=m\phi+g\phi^2$
with $m=10$ and $g=0,3$. We employed
a HMC algorithm \cite{hmc} in which
the fermions were replaced by (real) pseudofermions $\chi$ with
action
\begin{displaymath}
S_F=\chi (M^TM)^{-1} \chi
\end{displaymath}
We obtained substantial improvement by use of
Fourier acceleration techniques described in \cite{us1,alg}. Using
these ideas we amassed data for
lattices from $L=4$ through $L=32$ (1 million and $2\times 10^4$
trajectories respectively)
both for $g=0$ and $g=3$ \cite{us2}.
Table~\ref{table2} illustrates the mean action obtained at $g=3$ as
a function of lattice size
\begin{table}[htb]
\caption{Mean action vs L}
\label{table2}
\begin{tabular}{@{}ll}
\hline
$<S>$ & L \\
\hline
4  & 31.93(6) \\
8  & 127.97(7) \\
16 & 512.0(3)\\
32 & 2046(3)\\ 
\hline      	
\end{tabular}
\end{table}
By fitting the zero-momentum correlation functions we obtained the
following massgaps as a function of lattice spacing $a=1/L$.
\begin{table}[hb]
\caption{Boson and fermion massgaps vs L}
\label{table3}
\begin{tabular}{@{}lll}
\hline
Lattice size & $m_B$ & $m_F$ \\
\hline
4 &  5.09(2) & 4.95(8) \\
8 &  6.52(2) & 6.44(5) \\
16 & 7.76(4) & 7.75(6) \\
32 & 8.29(19) & 8.33(30) \\
\hline      	
\end{tabular}
\end{table}
Fig~\ref{fig3} shows a plot showing the
bosonic and fermionic contributions to the
Ward identity in
eqn.\ref{invariance}. If the Ward identity is satisfied these
two curves should sum to zero as the figure confirms. 
\section{Conclusion and Discussion}
We have shown that it is possible
to write down a 
local lattice action for the 2D Wess Zumino model with
extended ($N=2$) supersymmetry which admits an
exact, local supersymmetry. 
The
presence of an exact SUSY-like symmetry together with
the finiteness of the theory then guarantees that the
full $N=2$ supersymmetry will be regained in the continuum
limit without fine tuning.
We have tested these ideas
by explicit
dynamical fermion simulations.

\end{document}